\documentclass{aastex631}

\begin{document}

\title{Estimating Coronal Mass Ejection Mass and 
Kinetic Energy 
by Fusion of Multiple Deep-learning Models}

\correspondingauthor{Khalid A. Alobaid}
\email{kaa65@njit.edu}

\author{Khalid A. Alobaid}
\affiliation{Institute for Space Weather Sciences, New Jersey Institute of Technology, University Heights, Newark, NJ 07102, USA}
\affiliation{Department of Computer Science, New Jersey Institute of Technology, University Heights, Newark, NJ 07102, USA}
\affiliation{College of Applied Computer Sciences, King Saud University, Riyadh 11451, Saudi Arabia}

\author{Yasser Abduallah}
\affiliation{Institute for Space Weather Sciences, New Jersey Institute of Technology, University Heights, Newark, NJ 07102, USA}
\affiliation{Department of Computer Science, New Jersey Institute of Technology, University Heights, Newark, NJ 07102, USA}

\author{Jason T. L. Wang}
\affiliation{Institute for Space Weather Sciences, New Jersey Institute of Technology, University Heights, Newark, NJ 07102, USA}
\affiliation{Department of Computer Science, New Jersey Institute of Technology, University Heights, Newark, NJ 07102, USA}

\author{Haimin Wang}
\affiliation{Institute for Space Weather Sciences, New Jersey Institute of Technology, University Heights, Newark, NJ 07102, USA}
\affiliation{Center for Solar-Terrestrial Research, New Jersey Institute of Technology, University Heights, Newark, NJ 07102, USA}
\affiliation{Big Bear Solar Observatory, New Jersey Institute of Technology, 40386 North Shore Lane, Big Bear City, CA 92314, USA}

\author{Shen Fan}
\affiliation{Institute for Space Weather Sciences, New Jersey Institute of Technology, University Heights, Newark, NJ 07102, USA}
\affiliation{Department of Computer Science, New Jersey Institute of Technology, University Heights, Newark, NJ 07102, USA}

\author{Jialiang Li}
\affiliation{Institute for Space Weather Sciences, New Jersey Institute of Technology, University Heights, Newark, NJ 07102, USA}
\affiliation{Department of Computer Science, New Jersey Institute of Technology, University Heights, Newark, NJ 07102, USA}

\author{Huseyin Cavus}
\affiliation{Department of Physics, Canakkale Onsekiz Mart University, 17110 Canakkale, Turkey}
\affiliation{Harvard-Smithsonian Center for Astrophysics, 60 Garden Street, Cambridge, MA 02138, USA}

\author{Vasyl Yurchyshyn}
\affiliation{Big Bear Solar Observatory, New Jersey Institute of Technology, 40386 North Shore Lane, Big Bear City, CA 92314, USA}

\begin{abstract}
Coronal mass ejections (CMEs) are massive solar eruptions, which 
have a significant impact on Earth.
In this paper, we propose a new method, called DeepCME, to estimate two properties of CMEs, namely, CME mass and kinetic energy.
Being able to estimate these properties helps better understand CME dynamics.
Our study is based on the CME catalog maintained
at the Coordinated Data Analysis Workshops (CDAW) Data Center,
which contains all CMEs manually identified since 1996 
using the Large Angle and Spectrometric Coronagraph (LASCO) 
on board the Solar and Heliospheric Observatory (SOHO).
We use LASCO C2 data in the period between
January 1996 and December 2020 to train, validate and
test DeepCME through 10-fold cross validation.
The DeepCME method is a fusion of three deep learning models, 
including ResNet, InceptionNet, and InceptionResNet. 
Our fusion model extracts features from LASCO C2 images, effectively combining the learning capabilities of
the three component models to jointly estimate the mass and kinetic energy of CMEs.
Experimental results show that
the fusion model yields a mean relative error (MRE) of
0.013 (0.009, respectively)
compared to the MRE of
0.019 (0.017, respectively)
of the best component model
InceptionResNet (InceptionNet, respectively)
in estimating the CME mass (kinetic energy, respectively).
To our knowledge, this is the first time that deep learning
has been used for CME mass and kinetic energy estimations.

\end{abstract}


\section{Introduction} \label{sec:intro}
Coronal mass ejections (CMEs) are massive solar eruptions 
that release billions of tons of charged particles 
into space at high speeds
\citep{2000JGR...105.2375L,WH2012LRSP}. 
These energetic phenomena are of significant importance, 
as they have the potential to disrupt the Earth's geomagnetic field, 
resulting in geomagnetic storms that can damage satellites, 
communication systems, and power grids \citep{2004SpWea...2.2004B}.
It is crucial to understand and forecast the properties of CMEs
to mitigate their potential harmful impact on our technological infrastructure. 
The study of CMEs has evolved over the years
\citep[e.g.,][]{Gopal2005JGRA,2012hssr.book.....S,
Pal2018ApJ,
KLM2019SpWea,
2020SpWea..1802478U,
MDT2022A&A}.
Early work focused on identifying solar features responsible for CMEs, 
such as magnetic field configurations
and the presence of solar flares \citep{2012hssr.book.....S}.
Over time, researchers have developed more advanced techniques,
including machine learning and artificial intelligence, 
for CME analysis
\citep[e.g.,][]{BI2016ApJ,2018ApJ...855..109L, 2019ApJ...881...15W,  2020ApJ...890...12L,
AAW2022, Guastavino2023}.
Deep learning, a subfield of machine learning and artificial intelligence, 
is now an effective predictive tool in
solar physics
\citep{ARCC2023LRSP}. 

The mass and kinetic energy of CMEs
are important characteristics that help scientists understand the dynamics of CMEs \citep{2012ApJ...752...36C}. 
Determining the mass and kinetic energy of CMEs has been a long-standing topic in
heliophysics \citep{1979SoPh...61..201M, 1981SoPh...69..169P, 2012ApJ...752...36C, 2017ApJ...844...61D, 2021ApJ...906...46N}. Traditionally, CME mass is estimated through observations of white-light coronagraphs, which record the brightness of the ejected material as it scatters sunlight \citep{2012ApJ...752...36C}. 
When these brightness measurements are converted into mass estimates, 
researchers can calculate the kinetic energy of a CME.
For example,
\citet{2010ApJ...722.1522V} investigated the dependence of the solar cycle on CME mass and kinetic energy over a full solar cycle (1996-2009) using LASCO coronagraph data. 
The authors discovered a sudden reduction in CME mass in mid-2003 
and identified a 6-month periodicity in the ejected mass starting from 2003.
\citet{2012ApJ...752...36C} utilized STEREO COR1 and COR2 coronagraphs to estimate the mass of a CME on 12 December 2008, revealing that the CME's dynamics was influenced by magnetic forces at heliocentric distances of less than or equal to 7 solar radii and solar wind drag forces at distances more than or equal to 7 solar radii. 
In another study,
\citet{2021ApJ...906...46N} presented a method for estimating the mass of halo CMEs using synthetic CMEs.
The authors concluded that the halo CME mass might be underestimated
when only the observed CME region was considered. 

In this paper, we propose DeepCME, which is a
fusion of three deep learning models,
to estimate the CME mass and kinetic energy using SOHO LASCO C2 data.
The three deep learning models are
ResNet, InceptionNet, and InceptionResNet.
In Section \ref{sec:data}, we describe the data used in our study. 
Besides LASCO C2 images \citep{1995SoPh..162..357B}, we also use the CME catalog, 
which we refer to as the CDAW catalog, maintained at the Coordinated Data Analysis Workshops (CDAW) Data Center \citep{2004JGRA..109.7105Y,2009EM&P..104..295G}.
Section \ref{sec:methods} presents the architecture and configuration details of DeepCME. 
Section \ref{sec:results} reports the experimental results. 
Section \ref{sec:conc} presents a discussion and concludes the article. 

It should be pointed out that our objective is to understand whether machine learning can capture hidden relationships between LASCO C2 observations and CME properties (mass, kinetic energy, occurrence rate, as well as other attributes documented in the CDAW catalog such as angular width, acceleration, etc.). Our experimental results in Section \ref{sec:results} show that the proposed DeepCME model is capable of inferring the relationships between LASCO C2 images and two important CME properties (mass and kinetic energy). These results demonstrate that deep learning could be a useful tool for helping to better understand CME dynamics.
We note that the most recent available CME mass and kinetic energy information in the CDAW catalog is from December 2020. Since January 2021, this information has been absent.
DeepCME could be used to
estimate the missing mass and kinetic energy information in the CDAW catalog
from January 2021 to the present.
Furthermore, the input of the DeepCME tool is obtained from directly observed images, which are available near real-time.
Thus, the tool has the potential to contribute to near-real-time 
CME mass and kinetic energy predictions.
Our work presents the first step toward the application of deep learning models to the estimation of CME attributes. Additional efforts are needed to explore the use of machine learning to predict the other properties of CMEs.

\section{Data} \label{sec:data}
\begin{figure}
\centering
\includegraphics[width=\textwidth]{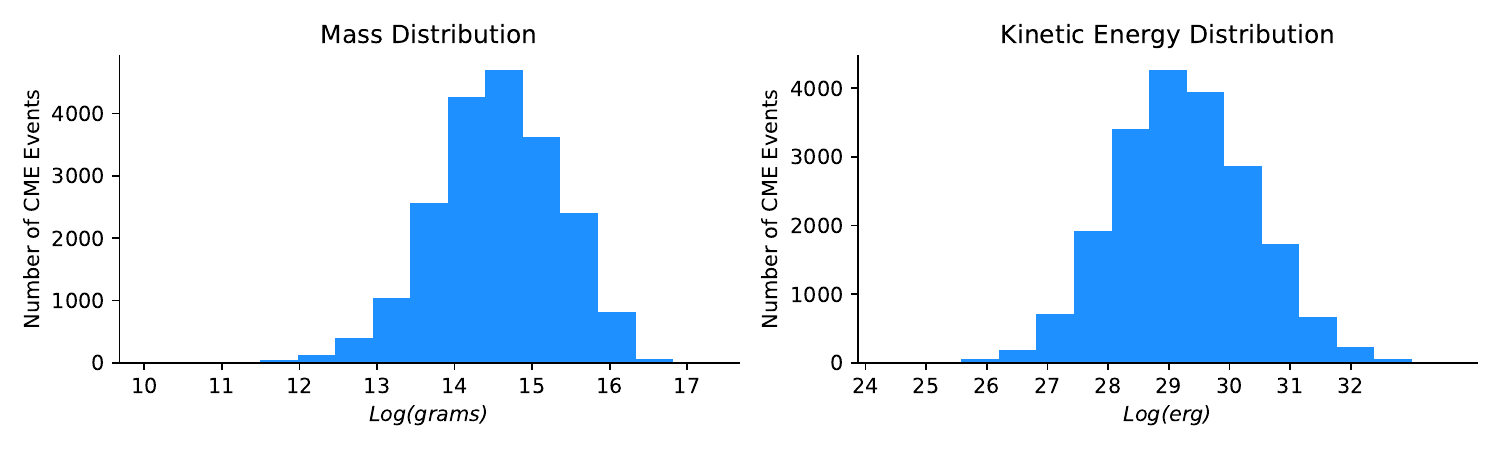}
\caption{Distributions of the mass and kinetic energy values
of the CMEs used in this study.}
\label{fig:fig_1}
\end{figure}

We start by collecting 20,084 CME events, spanning January 1996 to December 2020, 
from the CDAW catalog accessible at \url{https://cdaw.gsfc.nasa.gov/CME_list/}. 
The mass and kinetic energy values of the CME events range from $1.1\times10^{10}$ to $2.0\times10^{17}$ grams and
from $2.2\times10^{24}$ to $4.2\times10^{33}$ erg, respectively. 
Table \ref{table:y_data_statistics} shows the statistics of the data.
For example, the 25th percentile value $v$ in mass represents that
25\% of all mass values lie below $v$ 
and $(100 - 25)$\% = 75\% of all mass values lie above $v$.
The wide ranges of values shown in Table \ref{table:y_data_statistics} present a challenge to
a deep learning model,
as they could potentially hinder the model's ability to learn the underlying patterns effectively. 
To overcome this issue, we applied a common logarithmic transformation to the values of mass and kinetic energy.
This is a widely used technique to normalize data with large variations
\citep{2005ApJ...619.1160A,2005ApJ...619..599Y, 2010ApJ...722.1522V}. 
Figure \ref{fig:fig_1} shows the distributions of
the mass and kinetic energy values
after applying the logarithmic transformation. 
\begin{deluxetable*}{lrccccc}
\tablecaption{CME Mass and Kinetic Energy Statistics}
\tablewidth{0.8\textwidth}
\label{table:y_data_statistics}
\tablehead{
\colhead{Statistic} & \colhead{} & \colhead{Mass (grams)} & \colhead{} & \colhead{} & \colhead{Kinetic Energy (erg)} & \colhead{}
}
\startdata
Mean & & 1.496$\times 10^{15}$ & & & 4.746$\times 10^{30}$ & \\
Median & & 3.500$\times 10^{14}$ & & & 1.700$\times 10^{29}$ & \\
Minimum & & 1.100$\times 10^{10}$ & & & 2.200$\times 10^{24}$ & \\
25th Percentile & & 1.100$\times 10^{14}$ & & & 3.000$\times 10^{28}$ & \\
75th Percentile & & 1.300$\times 10^{15}$ & & & 1.000$\times 10^{30}$ & \\
Maximum & & 2.000$\times 10^{17}$ & & & 4.200$\times 10^{33}$ & \\
\enddata
\end{deluxetable*}

For each CME event,
we downloaded its corresponding LASCO C2 images \citep{1995SoPh..162..357B}
from the European Space Agency SOHO Science Archive (https://ssa.esac.esa.int/) utilizing the SunPy library \citep{2015CS&D....8a4009S}.
These images, with a size of $1024 \times 1024$, 
provide a comprehensive view of CMEs during their first appearance
at 1.5 solar radii
in the LASCO C2 field of view, 
allowing scientists to capture the initial characteristics of the events. 
To optimize computational efficiency, we resize the images from their original dimension to a size of
$256 \times 256$. 
To make data handling feasible and ensure a representative sample over years, we randomly selected 10\% CME events from each year. 
C2 images having multiple CME events were excluded from the study.

\begin{figure}
\centering
\includegraphics[width=0.7\textwidth]{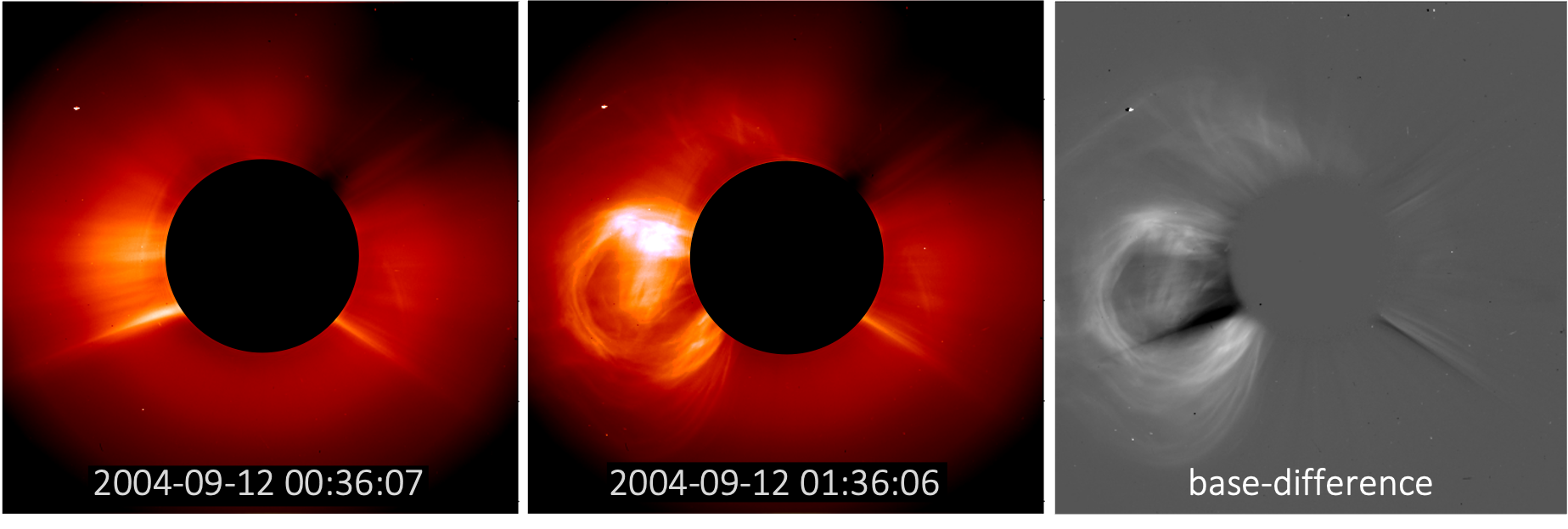}
\caption{Construction of the base-difference image for the CME event that occurred on 12 September 2004 at 00:36:06 UT. 
The left panel shows the pre-event image of the CME.
The middle panel shows 
the CME appearing as a full-grown structure.
The right panel shows the base-difference image of the CME
obtained by subtracting the image in the left panel from the image in the middle panel.}
\label{fig:base_difference}
\end{figure}

Following \citet{2019ApJ...881...15W},
for each selected CME event,
we constructed a base-difference image
by subtracting its pre-event image from 
the image in which the CME appears as a full-grown structure.
Here, ``full-grown'' refers to the last LASCO frame when
all three parts of the CME (i.e., its core, cavity, and leading edge \citep{2020JGRA..12528139B})
are visible within the field of view. 
A CME event without either the pre-event image or the image in which the CME appears as a full-grown structure was
excluded from the study.
Construction of this base-difference image
allows us to isolate and highlight the changes explicitly associated with the CME event.
Figure \ref{fig:base_difference} illustrates how a base-difference image
is constructed.

The above process resulted in a set of 1,964 base-difference images
corresponding to 1,964 selected CME events, where
each base-difference image uniquely represents a CME event.
For each selected CME event and its corresponding base-difference image, we used the common logarithm of its mass and kinetic energy, respectively,
as the ground-truth label for the event.
We adopt a 10-fold cross-validation scheme
in which the set of 1,964 images is
randomly partitioned into 10 subsets or folds of equal size. 
In the run $i$, the fold $i$ is used for testing, and
the union of the other nine folds is used for training.
10\% of the training set is used for validation.
There are 10 folds and, therefore, 10 runs.
The mean and standard deviation of the
predicted mass and kinetic energy values are calculated
over the 10 runs and plotted, respectively.

\section{Methodology} \label{sec:methods}
\subsection{Component Models}
To extract features from the 
base-difference images, 
we employ three
deep learning models:
ResNet50 \citep{DBLP:conf/cvpr/HeZRS16}, 
InceptionV3 \citep{DBLP:conf/cvpr/SzegedyVISW16}, 
and InceptionResNetV2 \citep{DBLP:conf/aaai/SzegedyIVA17}. 
The three deep learning models are among the
most widely used convolutional neural networks for computer vision applications. 
We also experimented with other classical models
such as
EfficientNet \citep{DBLP:conf/icml/TanL19} and 
VGGNet \citep{DBLP:journals/corr/SimonyanZ14a}, 
which yielded worse performance.

\color{black}{The ResNet50 model belongs to the class of residual networks \citep{DBLP:conf/cvpr/HeZRS16}.} 
It begins with a $7 \times 7$ convolutional layer with 64 filters
and a stride of 2, 
followed by a $3 \times 3$ max pooling layer with a stride of 2. 
Next, the model consists of four parts, 
each containing a sequence of residual blocks. 
These blocks, also known as bottleneck blocks, 
are the building blocks of the ResNet50 architecture \citep{DBLP:conf/cvpr/HeZRS16}. 
The InceptionV3 model begins with a $3 \times 3$ convolutional layer
with 32 filters and a stride of 2, 
followed by another $3 \times 3$ convolutional layer
with 32 filters and a stride of 1 \citep{DBLP:conf/cvpr/SzegedyVISW16}. 
This part is then followed by a $3 \times 3$ convolutional layer
with 64 filters and a stride of 1, and a $3 \times 3$ max pooling layer
with a stride of 2. 
Next, the model contains three inception modules, 
each with 288 filters, with a grid size of $35 \times 35$. 
This part is reduced to a $17 \times 17$ grid and
then to a $8 \times 8$ grid \citep{DBLP:conf/cvpr/SzegedyVISW16}. 
The InceptionResNet model introduces a simple yet
effective concept in which it combines the multi-scale feature learning of inception modules with the capabilities of ResNet's residual connections \citep{DBLP:conf/aaai/SzegedyIVA17}. 

The three component models were pre-trained on the ImageNet dataset
\citep{DBLP:conf/cvpr/DengDSLL009}, which contains 1,000 object classes with
approximately 1.2 million annotated images.
To adapt their architectures for the regression tasks of estimating CME mass and kinetic energy,
we modify each component model to suit our specific requirements
by removing its final fully connected layer
and activation function, as the regression tasks require continuous output values
instead of discrete class probabilities.

\subsection{The Fusion Model}
\begin{figure}
\centering
\includegraphics[width=\textwidth]{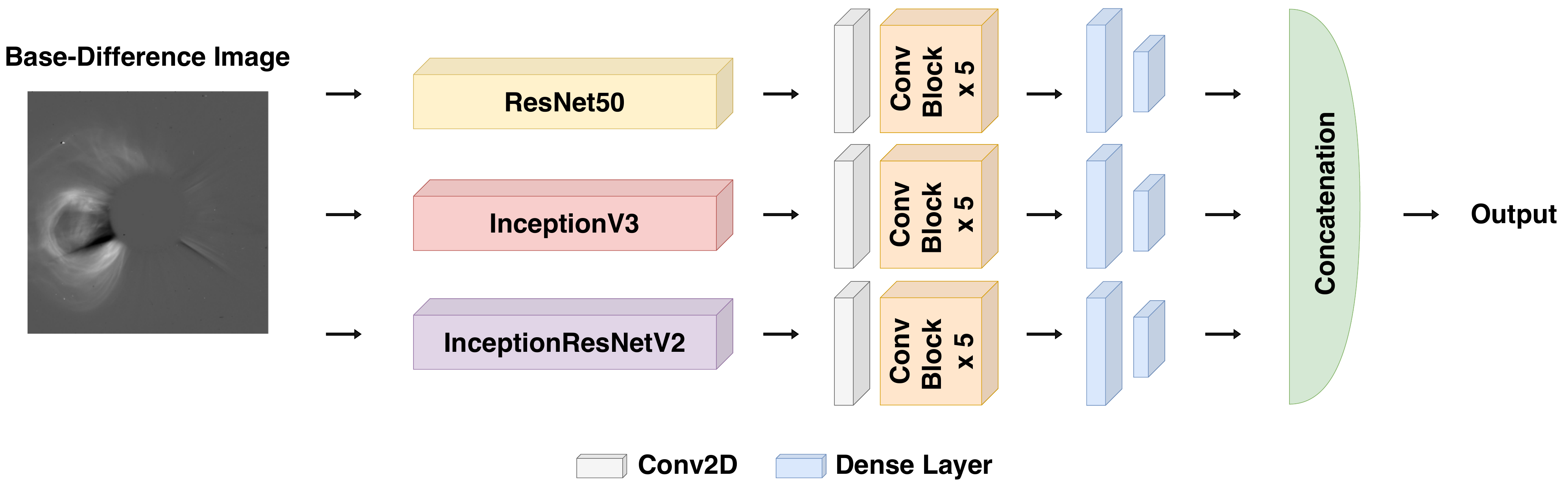}
\caption{Illustration of the DeepCME architecture. 
The fusion model begins with three component models,
namely ResNet50, InceptionV3, and InceptionResNetV2, 
each of which is
succeeded by a 2D convolutional layer, followed by five convolutional blocks, 
followed by two dense layers 
with 1024 neurons and 1 neuron, respectively.
The fusion model concludes with
a concatenation layer, which produces the output (i.e., the estimated CME mass or kinetic energy value) for the input base-difference image (CME event).}
\label{fig:DeepCME}
\end{figure}

DeepCME is a fusion of the three component models
described above.
Each input base-difference image, representing a CME event, is fed to
the component models, respectively.
Each component model is
succeeded by a two-dimensional (2D) convolutional layer, 
followed by five convolutional blocks. 
The last convolutional block is followed by
two dense layers, with 1024 neurons and 1 neuron, respectively.
Each component model pipeline predicts
an estimated value, respectively.
A concatenation layer then takes the median of the three
estimated values predicted by the three component model pipelines
to produce the final estimated value.
Figure \ref{fig:DeepCME} shows the architecture of
the DeepCME fusion model.
Table \ref{table:DeepCME_Architecture} presents the configuration
details of the fusion model.
\begin{deluxetable*}{lccccccc}
\tablecaption{Configuration Details of the DeepCME Model}
\tablewidth{0.8\textwidth}
\label{table:DeepCME_Architecture}
\tablehead{
\colhead{Layer} & \colhead{Type} & \colhead{Number of Filters} & \colhead{Kernel Size} & \colhead{Stride} & \colhead{Regularization} & \colhead{Activation} & \colhead{Output} 
}
\startdata
Conv2D & Convolutional & 64 & 11$\times$11 & 1 & - & LeakyReLU & 8$\times$8$\times$64\\
ConvBlock 1 & Convolutional & 64 & 11$\times$11 & 2 & Batch Norm & LeakyReLU & 4$\times$4$\times$64\\
ConvBlock 2 & Convolutional & 128 & 11$\times$11 & 1 & Batch Norm & LeakyReLU & 4$\times$4$\times$128\\
ConvBlock 3 & Convolutional & 128 & 11$\times$11 & 2 & Batch Norm & LeakyReLU & 2$\times$2$\times$128\\
ConvBlock 4 & Convolutional & 256 & 11$\times$11 & 1 & Batch Norm & LeakyReLU & 2$\times$2$\times$256\\
ConvBlock 5 & Convolutional & 256 & 11$\times$11 & 2 & Batch Norm & LeakyReLU & 1$\times$1$\times$256\\
Dense & Fully Connected & - & - & - & - & - & 1024 \\
Dense & Fully Connected & - & - & - & - & Linear & 1 \\
\enddata
\end{deluxetable*}

When estimating the CME mass, we feed all training base-difference images (training CME events) and
their corresponding ground-truth labels to DeepCME to train the fusion model.
The model is trained for 1000 epochs, with a batch size of 256.
We use the adaptive moment estimation optimizer (Adam) and the mean absolute error (MAE) as the loss function \citep{DBLP:journals/siamrev/Berk92}. 
Table \ref{table:DeepCME_Hyperparameters} summarizes
the hyperparameters used for DeepCME training.
During testing, we input each test base-difference image 
(test CME event) into the trained fusion model,
which predicts an estimated CME mass value for the test event.
Similarly, when the CME kinetic energy is estimated, 
we feed all training base-difference images (training CME events) and their
corresponding ground-truth labels to DeepCME to train the fusion model.
The hyperparameters used in the training are the same as those in
Table \ref{table:DeepCME_Hyperparameters}.
During testing, we input each test base-difference image 
(test CME event)
into the trained fusion model,
which predicts an estimated kinetic energy value for the test event.
\begin{deluxetable*}{ccccc}
\tablecaption{Hyperparameters for DeepCME Training}
\tablewidth{0.8\textwidth}
\label{table:DeepCME_Hyperparameters}
\tablehead{
 \colhead{Loss Function} & \colhead{Optimizer} & \colhead{Initial Learning Rate} & \colhead{Batch Size} & \colhead{Epoch}
}
\startdata
MAE & Adam & 0.001 & 256  & 1000\\
\enddata
\end{deluxetable*}

\newpage
\section{Results} \label{sec:results}
\subsection{Performance Metrics} \label{sec:setup_and_metrics}
We use four metrics to evaluate the performance of
the DeepCME fusion model and its component models.
These metrics include
the mean absolute error (MAE), 
the mean relative error (MRE), 
the coefficient of determination ($R$-squared or $R^2$), 
and Pearson's product-moment correlation coefficient 
\citep[PPMCC;][]{Pearson_PPMCC,DBLP:journals/siamrev/Berk92,Jiang2022ApJ}.
In what follows, 
$y_i$ denotes the true value 
of the $i$th base-difference image (CME event) in the test set,
$\hat{y}_i$ denotes the predicted value of the $i$th base-difference image (CME event) 
in the test set,
$n$ is the total number of base-difference images (CME events) in the test set, and
$\bar{y}$ = $\frac{1}{n}$$\sum_{i=1}^{n} y_{i}$ denotes the mean of the true values
for all base-difference images (CME events) in the test set.

The first metric is defined as
\begin{equation} 
\mbox{MAE} = \frac{1}{n} \sum_{i=1}^{n} |\hat{y}_i - y_i|, 
\end{equation}
\noindent
which calculates the average absolute difference
between the predicted value and the true value
\citep{DBLP:journals/siamrev/Berk92}. 
A smaller MAE signifies a better fit of a model to the data, 
implying the model's better predictive performance. 

The second metric is defined as
\begin{equation} 
\mbox{MRE} = \frac{1}{n} \sum_{i=1}^{n} \left|\frac{\hat{y}_i - y_i}{y_i}\right|, \end{equation}
\noindent
which calculates the average relative difference 
between the predicted value and the true value. 
A smaller MRE indicates better model performance. 

The third metric is defined as 
\begin{equation} 
R^2 = 1 - \frac{\sum_{i=1}^{n} (\hat{y}_i - y_i)^2}{\sum_{i=1}^{n} (y_i - \bar{y})^2}, 
\end{equation}
\noindent
which measures the strength of the
relationship between predicted and true values
in the test set. 
It ranges from
$-\infty$ to 1,  
with a higher value indicating better model performance. 

The fourth metric is defined as 
\begin{equation}
\mbox{PPMCC} = \frac{Exp[(X-\mu _X)(Y-\mu _Y)]}{\sigma _X \sigma_Y},
\end{equation}
\noindent
where $X$ and $Y$ represent the predicted values and true values, respectively;
$\mu_X$ and $\mu_Y$ are the mean of $X$ and $Y$, respectively; 
$\sigma_X$ and $\sigma_Y$ are the standard deviation of $X$ and $Y$ respectively; 
and $Exp(\cdot)$ stands for the expected value.
PPMCC measures the linear correlation between predicted and true values in the test set \citep{Pearson_PPMCC}. 
It ranges from $-1$ to 1, with $-1$ indicating a perfect negative correlation, 
1 representing a perfect positive correlation and 0 meaning that there is no correlation.

\subsection{Performance Evaluation}

\begin{figure}
\centering
\includegraphics[width=\textwidth]{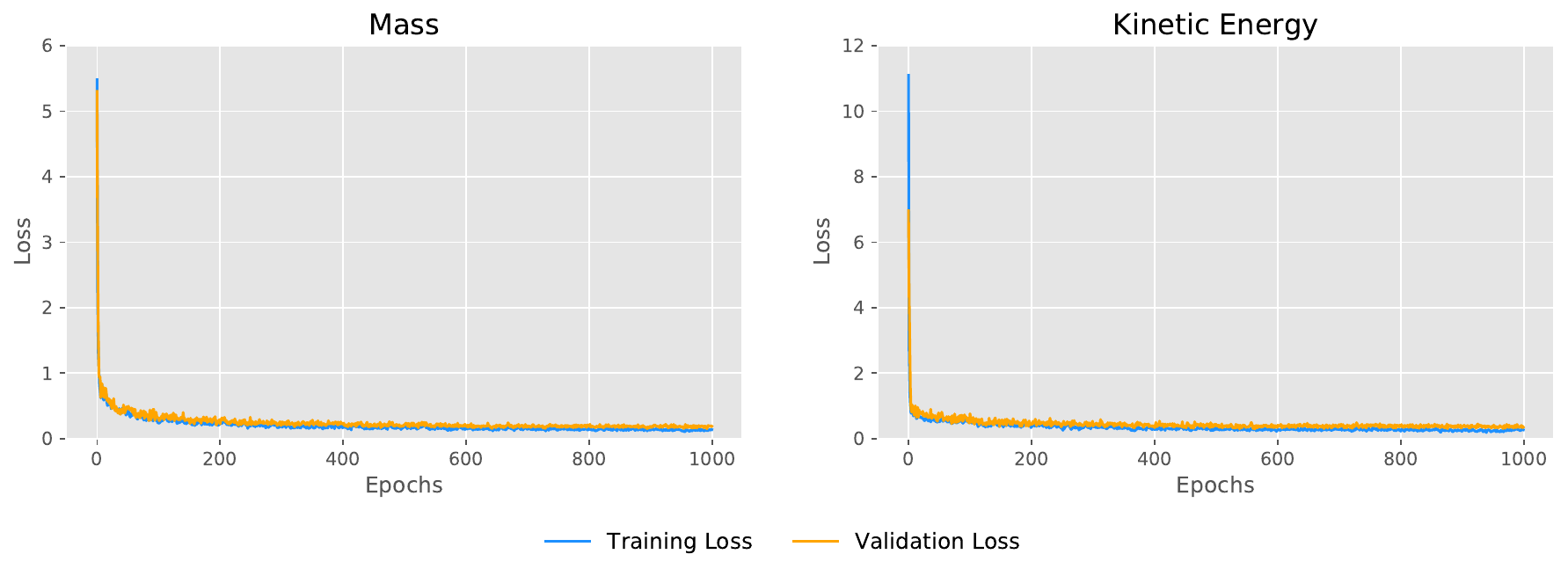}
\caption{Training and validation learning curves showing 
DeepCME is a well-fit model in estimating the mass and kinetic energy, respectively, of CMEs.}
\label{fig:learning_curve}
\end{figure}

We conducted a series of experiments to
understand the behavior of DeepCME
and evaluate the performance of DeepCME and its three component models (ResNet50, InceptionV3, and InceptionResNetV2). 
The evaluation was carried out using the 10-fold cross-validation scheme
described in Section \ref{sec:data},

which is a standard technique to detect overfitting.
Figure \ref{fig:learning_curve} presents the DeepCME training and validation learning curves.
The downward and convergence trends in the learning curves demonstrate DeepCME's ability to learn and generalize well to unseen data, with a decrease in the training loss and validation loss, respectively, as the number of epochs increases.
The learning curves in Figure \ref{fig:learning_curve}
show that DeepCME is a well-fit model.

\begin{figure}
\centering
\includegraphics[width=\textwidth]{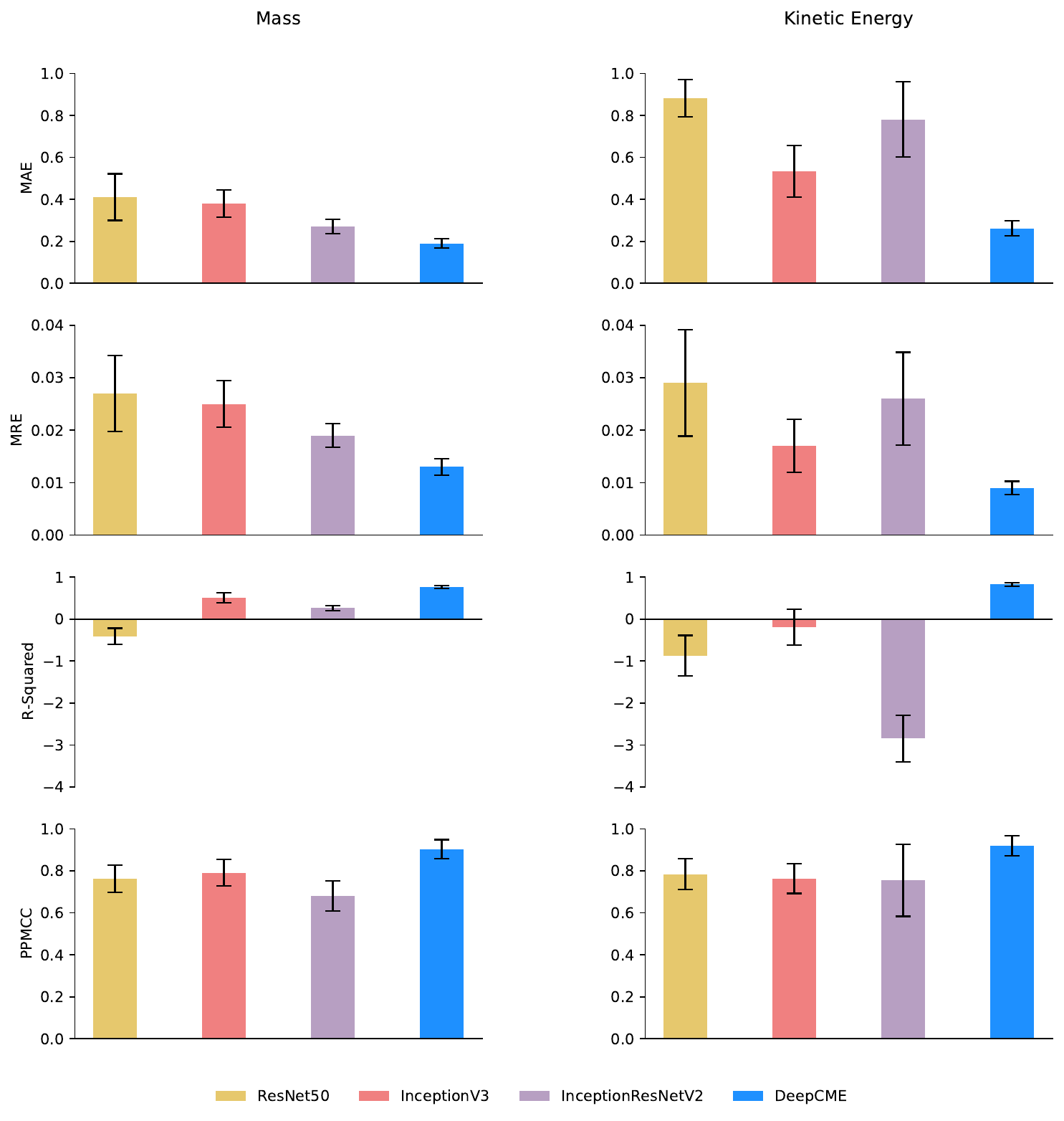}
\caption{Comparison between DeepCME and its three component models.
Left column: performance metric values, displayed by bar charts, obtained by the four models in estimating the mass of CMEs.
Right column: performance metric values obtained by the four models in estimating the kinetic energy of CMEs.}
\label{fig:barchart}
\end{figure}

Figure \ref{fig:barchart} compares DeepCME 
with its three component models.
In the figure, each colored bar represents the mean of the 10 runs in cross-validation, and its associated
error bar represents the standard deviation
divided by the square root of the number of runs
\citep{AAW2022,2022XGBoostSYMHbyIong}.
When estimating CME mass, the DeepCME model
performs better than the other three models,
as shown in the left column of Figure \ref{fig:barchart}.
DeepCME produces the lowest MAE of 0.190,
the lowest MRE of 0.013,
the highest $R^2$ of 0.763,
and the highest PPMCC of 0.904.
The InceptionV3 model achieves the second best $R^2$ of 0.505 and the PPMCC value of 0.791. The InceptionResNetV2 model ranks second in MAE and MRE with 0.271 and 0.019, respectively.
When estimating the kinetic energy of CMEs,
the DeepCME model also outperforms the other three models,
as shown in the right column of Figure \ref{fig:barchart}.
DeepCME achieves the lowest MAE of 0.262, the lowest MRE of 0.009, the highest $R^2$ of 0.828, and the highest PPMCC of 0.920. 
The InceptionV3 model ranks second on MAE, MRE and $R^2$ with 0.534, 0.017, and $-0.19$, respectively. ResNet50 is the second best model in PPMCC with a value of 0.784.
Furthermore, DeepCME has the smallest standard deviation and exhibits the most stable behavior
among the four models.
This happens because DeepCME works by taking the median of
the values predicted by the three component model pipelines,
resulting in smoother results than the individual component models.

\begin{figure}
\centering
\includegraphics[width=\textwidth]{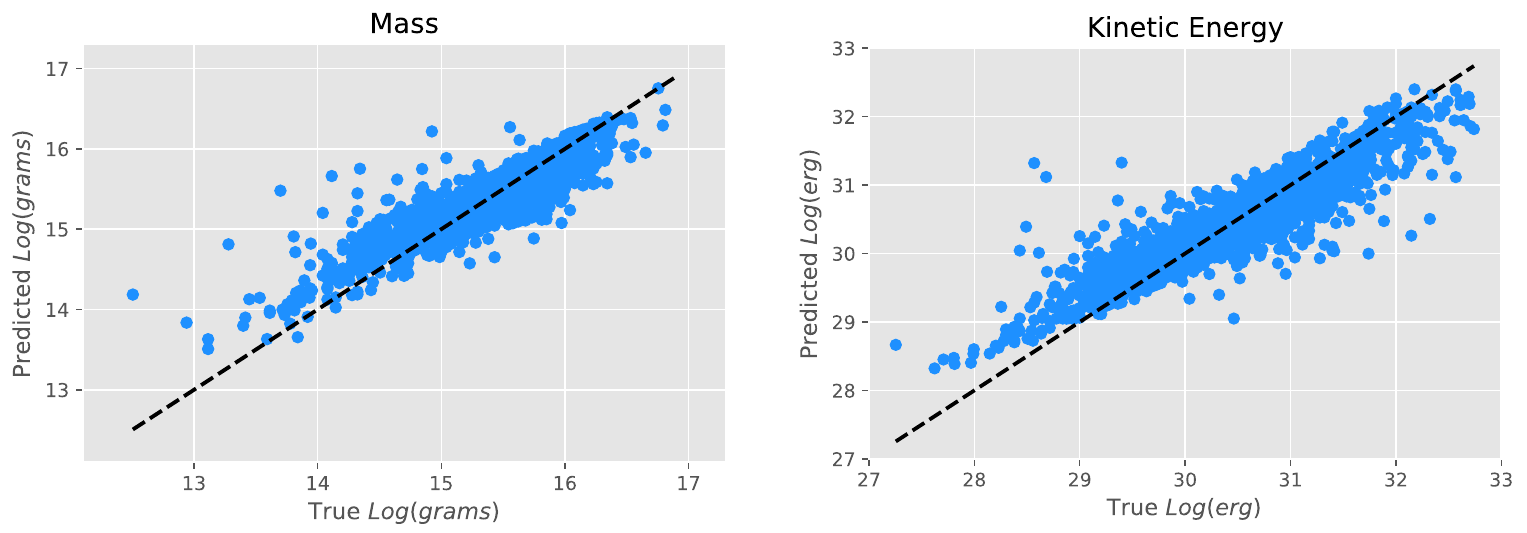}
\caption{Scatter plots showing DeepCME's predicted values
versus ground truth values in
estimating the mass and kinetic energy, respectively, of CMEs.}
\label{fig:scatterplot}
\end{figure}

\begin{figure}
\centering
\includegraphics[width=\textwidth]{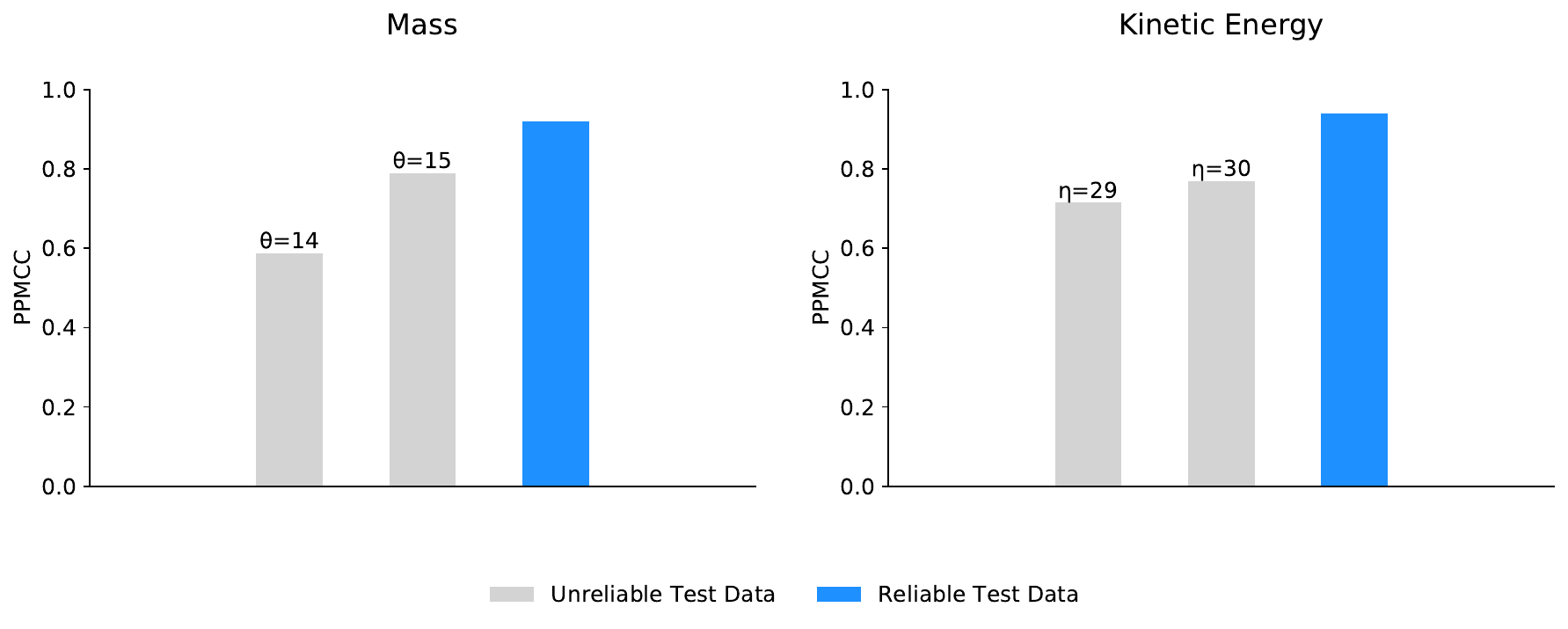}

\caption{Reliability assessment of DeepCME showing the performance
of the model on reliable test data and unreliable test data respectively.}
\label{fig:reliability}
\end{figure}

Figure \ref{fig:scatterplot} presents scatter plots
that visualize the relationship between the predicted values of DeepCME and the actual values
when estimating the CME mass and kinetic energy, respectively. 
The X axis denotes the ground truth values,
while the Y axis denotes the predicted values. 
It can be seen from Figure \ref{fig:scatterplot} that
the low mass/kinetic energy predictions deviate more than the high mass/kinetic energy predictions.
This happens because there are fewer CMEs with low mass/kinetic energy (see Figure \ref{fig:fig_1}), 
and consequently, DeepCME does not acquire enough knowledge during training to make accurate predictions on them.
We further conducted a reliability assessment
of DeepCME
by dividing the test data into reliable test data and unreliable test data \citep{DBLP:journals/jbi/NicoraRAB22}.
When estimating the mass of CMEs, reliable test data contain CMEs whose mass values range
from 15 to 17 log(grams), and unreliable test data contain CMEs
whose mass values are less than a threshold, $\theta$ log(grams),
where $\theta$ is 14 and 15 respectively.
When estimating the kinetic energy of CMEs, 
reliable test data contain CMEs whose kinetic energy values
range from 30 to 33 log(erg), and unreliable test data contain CMEs
whose kinetic energy values are less than a threshold, $\eta$ log(erg),
where $\eta$ is 29 and 30 respectively.
Figure \ref{fig:reliability} compares the PPMCC values obtained
by running DeepCME on reliable test data and
unreliable test data, respectively.
It can be seen in Figure \ref{fig:reliability} that predictions with lower mass/kinetic energy values are less reliable (with smaller PPMCC values) 
than predictions with higher mass/kinetic energy values, 
a finding consistent with the scatter plots presented in Figure \ref{fig:scatterplot}.

\color{black}

\section{Discussion and Conclusion} \label{sec:conc}
We present DeepCME, a deep learning fusion model designed to estimate the mass and kinetic energy
of a CME in the CDAW catalog
given the LASCO C2 base-difference image that uniquely represents the event. 
DeepCME combines the strengths of three component models
(ResNet, InceptionNet, and InceptionResNet) to extract features from
the base-difference images of CME events and to make predictions.
Experimental results
based on data from January 1996 to December 2020
using a 10-fold cross-validation scheme
demonstrate the good performance of DeepCME. 
The fusion model yields a mean relative error (MRE) of 
0.013 (0.009, respectively)
compared to the MRE of 
0.019 (0.017, respectively)
of the best component model
InceptionResNet (InceptionNet, respectively)
in estimating the CME mass (kinetic energy, respectively).

We have used LASCO C2 level 0.5 images in our work. 
In separate experiments, we adopted level 1.0 images to 
train and test DeepCME.
The level 0.5 images are raw data, while the level 1.0 images are calibrated data. 
Our results show that
there is not much difference between the
level 0.5 images and the level 1.0 images
in terms of prediction accuracy.
This happens probably because operations such as image flipping and image warping
in the calibration process
have no impact on a machine learning system.

\color{black}{In the study presented here, 
we used a base-difference image to uniquely represent a CME event.}
In an additional experiment, we explored an alternative approach in which
we used a complete set of LASCO C2 images
that spanned a time frame of 10 minutes before 
and up to 2 hours 
after the onset time of 
a CME to represent the CME event \citep{2019ApJ...881...15W}.
All the C2 images shared the same ground-truth label, 
i.e. the common logarithm of
the mass and kinetic energy, respectively, of the event.
The results obtained from this experiment indicate that the use of complete sets of images leads to worse performance than the use of unique base-difference images.
Specifically, DeepCME yields a mean relative error (MRE) of
0.024 (0.021, respectively) when using complete sets of images compared to the MRE of 0.013 (0.009, respectively) 
obtained by using
unique base-difference images
in estimating the CME mass (kinetic energy, respectively).
In theory, one would need to label the different images of a CME event with different kinetic energy values
while taking into account the velocity of the CME.
However, the CDAW catalog provides only one kinetic energy value for each CME event,
rather than one kinetic energy value for each image.
Assigning the same ground-truth label to different images of a CME event
would confuse a machine learning model, which would yield worse performance.
We conclude that the proposed approach of using unique base-difference images is a viable one for
CME mass and kinetic energy estimations.

\begin{acknowledgments}
We appreciate the anonymous referee for constructive comments and suggestions.
We thank members of the Institute for Space Weather Sciences for fruitful discussions.
K.A. is supported by King Saud University, Saudi Arabia.
J.W. and H.W. acknowledge support from NSF grants AGS-1927578, AGS-2149748, AGS-2228996 
and OAC-2320147.
H.C. is supported by the Fulbright Visiting Scholar Program of the Turkish Fulbright Commission.
V.Y. is supported by the NSF grant AGS-2300341.
The CME catalog used in this work was created and maintained at the CDAW Data Center by NASA and the Catholic University of America in cooperation with the Naval Research Laboratory. 
SOHO is an international cooperation project between ESA and NASA. 
\end{acknowledgments}

\bibliographystyle{aasjournal}

\end{document}